\def\frac#1#2{{#1 \over #2}}

\def\half{\ifinner {\scriptstyle {1 \over 2}}
   \else {1 \over 2} \fi}

\def\bra#1{\langle#1\vert}  
\def\ket#1{\vert#1\rangle}  

\def\tr{\mathop{\rm tr}\nolimits} 
\def\Sp{\mathop{\rm Sp}\nolimits} 

\def\bm#1{\mbox{\boldmath$#1$}}

\def\fpi{f_\pi}
\def\mpi{m_\pi}
\def\bra#1{\langle #1 \,\vert}
\def\ket#1{\vert\, #1 \rangle}

\def\wlog#1{} 
\catcode`\@=11

\def\wlog{\immediate\write\m@ne} 
\catcode`\@=12 
\documentstyle[preprint,aps,12pt]{revtex}
\begin{document}
\date{\today}
\listoffigures
\title{$1/N_c$ Rotational Corrections to $g_A$ and\\
Isovector Magnetic Moment of the Nucleon}
\author{Chr.V.Christov$^{1)}$\thanks{
Permanent address:Institute for Nuclear Research and Nuclear Energy, Sofia,
Bulgaria}, K.Goeke${^1)}$, P.Pobilitsa$^{1,2)}$, V.Petrov$^{2)}$,\\
M.Wakamatsu$^{1,3)}$ and T.Watabe$^{1,4)}$}
\address{
$^{1)}$Institut f\"ur Theoretische Physik II, Ruhr-Universit\"at Bochum,
D-44780 Bochum, Germany\\
$^{2)}$St.Petersburg Nuclear Physics Institute, Gatchina,
St.Petersburg 188350,Russia\\
$^{3)}$Department of Physics, Faculty of Science, Osaka University,
Toyonaka, Osaka 560, Japan\\
$^{4)}$Department of Physics, Tokyo Metropolitan University, Hachiohji,
Tokyo 192, Japan}
\maketitle

\begin{abstract}
{The $1/N_c$ rotational corrections to the axial vector constant
and the isovector magnetic moment of the nucleon are studied in the
Nambu -- Jona-Lasinio model. We follow a semiclassical quantization
procedure in terms of path integrals in which we can include
perturbatively corrections in powers of angular
velocity $\Omega \sim \frac 1{N_c}$. We find non-zero $1/N_c$ order
corrections from both the valence and the Dirac sea quarks. These
corrections are large enough to resolve the long-standing problem of a
strong underestimation of both $g_A$ and $\mu^{IV}$ in the leading
order. The axial constant $g_A$ is well reproduced, whereas the
isovector magnetic moment $\mu^{IV}$ is still underestimated by 25 \%.}
\end{abstract}
\newpage  
Among the effective low-energy models of baryons the chiral soliton model,
based on a semibosonized version~\cite{Eguchi76} of the Nambu --
Jona-Lasinio lagrangean~\cite{Nambu61}, seems to play an essential role.
In fact, an equivalent effective chiral quark-meson lagrangean can be
derived~\cite{Diakonov84} from the instanton model of the QCD vacuum.
This chiral soliton model, which is frequently referred simply as NJL model,
incorporates the general accepted phenomenological
picture of the baryon as a bound state of $N_c$ valence quarks coupled to the
polarized Dirac sea ($\overline qq$ pairs). Operationally in the model the
nucleon problem is solved in two steps~\cite{Diakonov89}. In the first
step, motivated by the large $N_c$ (number of colors) limit, a static
localized solution (soliton) of a hedgehog structure is found. This
hedgehog solution does not preserve the spin and isospin. Making use
of the rotational zero modes in a semiclassical quantization scheme one can
assign proper spin and isospin quantum numbers to the soliton.
The model is rather successful in describing both the static nucleon
properties~\cite{Diakonov89,Reinhardt89W,TMeissner89,Goeke91,Wakamatsu91}
and the nucleon form factors~\cite{TMeissner91,Gorski92}. The only
exceptions are the axial  coupling constant
$g_A$ and the isovector magnetic moment $\mu^{IV}$, which are strongly
underestimated ($g_A\approx 0.8$ and $\mu^{IV}\approx 2.4$(n.m.)) in the
leading order, compared to the experimental values of 1.25 and 4.71 (n.m.).
This is not a problem just of the NJL model. In
fact, most of the chiral models, if they assume physical values of the
pion decay constant $\fpi$ and the pion mass $\mpi$, yield rather
strong deviations from the experimental values for these quantities.
In the case of the NJL model the simplest solution to this problem is
to take into account the rotational
corrections coming from the next to leading order terms.

In a very recent paper Wakamatsu and Watabe~\cite{Wakamatsu93}
estimated in the NJL model $1/N_c$ rotational corrections to $g_A$ thus
providing a step in the right direction. They found a considerable non-zero
valence contribution ($\approx 0.4$). In this work they made the important
observation that in the case of $g_A$, after the canonical
quantization, the collective operators do not commute. However, in their
scheme the non-zero result is due to a particular order of the
collective operators being not fully justified by path integrals or
many-body techniques. As a consequence
the valence contribution includes transitions between occupied levels,
which violate the Pauli principle (even though, numerically, this Pauli
violating contribution turns out to be only a tiny fraction of the valence
contribution), and the Dirac sea
$1/N_c$ contribution vanishes exactly. The latter is particularly puzzling
since apart from the regularization there is no any principle
difference between the valence and the sea quarks in the NJL model.
Based on the formulae of Wakamatsu and Watabe~\cite{Wakamatsu93}
Alkofer and Weigel~\cite{Alkofer93} studied the axial coupling constant
in the context of the PCAC. From their values for $g_A$ one can
estimate how far the PCAC holds up in the linear order in $\Omega$ and
in particular, a violation of less than 2 \% for a reasonable value  of
the constituent mass $M=400$ MeV can be guessed.

Obviously a satisfying theoretical and numerical treatment of higher
order rotational corrections in the semiclassical quantization scheme
is still missing. It is therefore the objective of the present work to
evaluate the $1/N_c$ rotational corrections to both the axial  coupling
constant and the isovector magnetic moment in the NJL model. To that
end we will follow the theoretical scheme of Diakonov et
al.~\cite{Diakonov89} elaborated in terms of path integrals.

We start with the definitions of axial and isovector magnetic currents
of a fermion field $\Psi(x)$, $A_k^a(x)=\Psi^+(x)\gamma_0
\gamma_k\gamma_5\frac {\tau^a}2 \Psi(x)$ and $J_k^a
(x)=\Psi^+(x)\gamma_0\gamma_k \frac {\tau^a}2 \Psi(x)$,
respectively. Here $k$ means space components and $a$ stands for the
isospin index. We express the nucleon matrix element of a current
operator $\hat A_k^a$ as a path integral including quark
$\Psi,\Psi^+$ and meson fields $U$ in Minkowski space:
\begin{eqnarray}
&\bra{N({\bf p\prime})}\Psi^+(0)\hat O_k^a\Psi(0)\ket{N({\bf p})}
\mathop{=}\limits_{T\to-i\infty} \frac 1Z\int d^3xd^3ye^{-i{\bf
p^\prime\cdot x}}e^{i{\bf p\cdot y}}\nonumber\\
&\times\int{\cal D}U\int {\cal D}\Psi {\cal D}\Psi^+
J_N(T/2,{\bf x})J_N^+(-T/2,{\bf y}) \Psi^+(0)\hat O_k^a\Psi(0)
e^{\textstyle i\int d^4z\Psi^+D(U)\Psi} \label{Eq2}
\end{eqnarray}
The equality (\ref{Eq2}) should be understood as a limit at large
euclidean time separation. Here $Z$ is the normalization factor which
is related to the same path integral but without the current operator
$\Psi^+\hat O_k^a\Psi$, and $\hat O_k^a$ stands
for the matrix part $\gamma_0\gamma_k(\gamma_5)\frac {\tau^a}2$ of the
current. The Dirac operator $D(U)=i\partial_t-h(U)$ includes the
single-particle hamiltonian $h(U)=\frac
{\bm{\alpha}\cdot\bm{\nabla}}i+M\beta U^{\gamma_5}+m_0\beta$ with meson
fields $U^{\gamma_5}=e^{\textstyle i\gamma_5\bm{\tau}\cdot\bm{\pi}}$.
Here $\bm{\alpha}$ and $\beta$ are the Dirac matrices and $m_0$ being
the current quark mass. The composite operator:
\begin{equation}
J_N(x)=\frac
1{N_c!}\varepsilon^{\beta_1\cdots\beta_{N_c}}\Gamma^{\{f_1\cdots
f_{N_c}\}}_{JJ_3,TT_3}
\Psi_{\beta_1f_1}(x)\cdots\Psi_{\beta_{N_c}f_{N_c}}(x),
\label{Eq6} \end{equation}
carries the quantum numbers $JJ_3,TT_3$ (spin, isospin) of the nucleon,
where $\beta_i$ is the color index, and $\Gamma^{f_1\cdots f_{N_c}}_{
JJ_3,TT_3}$ is a symmetric matrix in flavor and spin indices $f_i$.

In eq.(\ref{Eq2}) we can integrate the quarks out:
\begin{eqnarray}
&\bra{N({\bf p^\prime})}\Psi^+(0)\hat O_k^a\Psi(0)\ket{N({\bf p})}=\frac 1Z
\Gamma^{\{g\}}_{JJ_3,TT_3}\Gamma^{\{f\}}_{JJ_3,TT_3}N_c \int d^3x d^3y
e^{-i{\bf p^\prime\cdot x}}e^{i{\bf p\cdot y}} \int{\cal D}U\nonumber\\
&\times\bigl\{\bra{T/2,{\bf x}}\frac iD \ket{0,0}_{f_1f^\prime}(\hat
O_k^a)_{f^\prime g^\prime}\bra{0,0} \frac iD \ket{-T/2,{\bf
y}}_{g^\prime g_1}-\Sp(\bra{0,0}\frac iD\ket{0,0}\hat O_k^a)\nonumber\\
&\times\bra{T/2,{\bf x}}\frac iD\ket{-T/2,{\bf y}}_{f_1g_1}\bigr\}
\prod\limits_{i=2}^{N_c}\bra{T/2,{\bf x}}\frac
iD\ket{-T/2,{\bf y}}_{f_ig_i}e^{\textstyle\tr\log D(U)}.
\label{Eq6a} \end{eqnarray}
In a natural way the result is split in a valence -- the first term, and
a Dirac sea contribution -- the second one (see the diagrams on the
l.h.s. of Fig.\ref{Figr1} a) and b)).

In order to integrate over the meson fields $U$  we start from a
stationary meson configuration of hedgehog structure $\bar
U(x)=e^{\textstyle i\bm{\tau}\cdot\hat{\bf x}\pi(x)}$
which minimizes the effective action. Then the integration over the meson
fields $U$ in the path integral can be done in a saddle point
approximation, which is motivated by the large $N_c$ limit.
In the next step we should allow the system to
fluctuate around the static hedgehog solution $\bar U(x)$ making use of the
rotational zero modes. Since the fluctuations which correspond to the zero
modes are not small they have to be treated ``exactly'' in the meaning of
path integral. Operationally it can be done introducing a rotating meson
fields of the form $U({\bf x},t)=R(t)\bar U({\bf x}) R^+(t)$,
where $R(t)$ is a time-dependent rotation SU(2) matrix in the isospin
space. It is easy to see that for such an ansatz one can
transform the effective action
$\quad\tr\log D(U)=\tr\log(D(\bar U)-\Omega)\quad$
as well as the quark propagator in the background meson fields $U$
\begin{equation}
\bra{T/2,{\bf x}}\frac i{D(U)} \ket{-T/2,{\bf y}}=R(T/2)\bra{T/2,{\bf
x}}\frac i{D(\bar U)-\Omega} \ket{-T/2,{\bf y}}R^+(-T/2),
\label{Eq10b}\end{equation}
where $\Omega=-iR^+(t)\dot R(t)=\frac 12\Omega_a\tau_a$
is the angular velocity matrix. Since $\Omega\sim \frac 1{N_c}$
(as can be seen below) one can consider $\Omega$ as a perturbation and
evaluate any observable as a perturbation series in $\Omega$ which is
actually an expansion in $\frac 1{N_c}$.

In this scheme the matrix element (eq.\ref{Eq2}) of the
current can be written as
\begin{eqnarray}
&\bra{N({\bf p\prime})}\Psi^+(0)\hat O_k^a\Psi(0)\ket{N({\bf p})}=\frac 1Z
N_c\int
d^3xd^3y d^3ze^{-i{\bf p^\prime\cdot x}}e^{i{\bf p\cdot y}} e^{i{\bf
(p^\prime-p)\cdot z}}
\int{\cal D}R \nonumber\\
&\times{D^J_{-T_3J_3}}^*(R(T/2))
\bigl\{\bra{T/2,{\bf x}}\frac i{D-\Omega}\ket{0,{\bf z}}R^+(0) \hat
O_k^a R(0) \bra{0,{\bf z}}\frac i{D-\Omega} \ket{-T/2,{\bf y}}\nonumber\\
&-\Sp(\bra{0,{\bf z}}\frac i{D-\Omega}\ket{0,{\bf z}}R^+(0)\hat
O_k^aR(0))\bra{T/2,{\bf x}}\frac i{D-\Omega} \ket{-T/2,{\bf
y}}\bigr\}\nonumber\\
&\times\bra{T/2,{\bf x}}\frac i{D-\Omega}\ket{-T/2,{\bf y}}^{\textstyle 2}
D^J_{-T_3J_3}(R(-T/2)) e^{\textstyle\tr\log(D-\Omega)}.
\label{Eq12} \end{eqnarray}
Here, the finite rotation matrix $D^J_{-T_3J_3}$, which carries the spin and
isospin quantum numbers of the nucleon, appears due to the rotations $R(t)$
of the valence quark propagators in eq.(\ref{Eq10b}) correlated by the
$\Gamma^{\{g\}}_{JJ_3,TT_3}$ matrices and the integral over ${\bf z}$ is due
to the translational zero modes treated in the leading order.

Now we are ready to make an expansion in $\Omega$. For the effective
action, it is well-known
procedure~\cite{Diakonov89,Goeke91,Wakamatsu91,Gorski92}, which
yields up to the second order in $\Omega$: \begin{equation}
\tr\log{(D-\Omega)}\approx \tr\log{D}+i\frac {\Theta}2\int dt\Omega_a^2.
\label{Eq14} \end{equation}
Here $\Theta$ is the moment of inertia. The first term will be absorbed in
$Z$ whereas the second one gives the evolution operator acting in the
space of matrix $R$. Expanding the quark propagator
$$
\frac 1{D-\Omega}\longrightarrow\frac 1D +\frac 1D\Omega\frac 1D+...
$$
we can separate the zero order ($\sim N_c^0$) and the linear
order ($\sim \frac 1{N_c}$) corrections in $\Omega$.
The expansion in  $\Omega$ is illustrated in Fig.\ref{Figr1} a) and  b) for
the valence contribution and for the Dirac sea one, respectively.

Henceforward we will concentrate on the linear order terms.
In this case we are left with the following path integral over $R$:
\begin{equation}
\int\limits_{R(-T/2)}^{R(T/2)}{\cal
D}R{D^J_{-T_3J_3}}^*(R(T/2))\frac 12\Sp(R^+(0)\tau^a
R(0)\tau^b)\Omega_c(t) D^J_{-T_3J_3}(R(-T/2))
e^{\textstyle i\frac{\Theta}2\int dt\Omega_c^2}. \label{Eq16}
\end{equation}
Here we use the identity $\quad (R^+(0)\hat O_k^a R(0))_{fg}=\frac
12\Sp(R^+(0)\tau^a R(0)\tau^b)(\hat O_k^b)_{fg}\quad$
in order to separate the $R(t)$ dependent part of the current which
does not carry flavor indices $fg$. The path integral (\ref{Eq16}) can
be taken~\cite{Diakonov89} rigorously within the approximation
(\ref{Eq14}). We obtain the well-known canonical quantization rule
$\quad\Omega_c\rightarrow J_c/\Theta$, where $J_a$ is the spin
operator. The final result for path integral (\ref{Eq16}) is a time ordered
product:
\begin{equation}
\vartheta(-t)D_{ab}(R(0))J_c+\vartheta(t)J_c D_{ab}(R(0)).
\label{Eq19} \end{equation}
which should be sandwiched between the nucleon rotational wave
function. In order to obtain the result (\ref{Eq19}) we essentially made use
of the basic feature of the path integral:
$$
\int\limits_{q_1=q(T_1)}^{q_2=q(T_2)}{\cal D}q
F_1(q(t_1))\cdots F_n(q(t_n))e^{iS}=\bra{q_2,T_2}T\{\hat
F_1(q(t_1))\cdots\hat F_n(q(t_n)\}\ket{q_1,T_1},
$$
namely that the path integral can be equivalently written as the
expectation value of the time ordered product of the corresponding
operators. Using (\ref{Eq19})
and the standard spectral representation of the quark propagator it is
straightforward to evaluate the matrix element of the current
eq.(\ref{Eq12}). In particular for the linear correction we get
$$
\bra{N(p^\prime)}\Psi^+(0)\hat O_k^a\Psi(0)\ket{N(p)}^{\Omega^1}=
\bra{J,J_3T_3}[\frac {J_c}\Theta,D_{ab}]\ket{J,J_3T_3}
$$
\vskip-0.5cm
\begin{equation}
\times N_c\sum\limits_{{n > val}\atop{m\leq val}}\frac
1{\epsilon_n-\epsilon_m}\bra{m}\tau_a\ket{n}\int d^3ze^{i{\bf
(p^\prime-p)\cdot z}}\Phi^+_n({\bf z})\hat O_k^b\Phi_m({\bf z}).
\label{Eq20}\end{equation}
Here $\Phi_n$ and $\epsilon_n$ are the eigenfunctions and eigenvalues
of the single-particle hamiltonian $h$. Since the $\tau$-matrix
element is asymmetric with respect to exchange of the states $m$ and $n$
a commutator appears in the collective matrix
element of eq.(\ref{Eq20}). It can be easily calculated:
\begin{equation}
\bra{J=1/2,J_3T_3}[\frac {J_c}\Theta,D_{ab}]\ket{J=1/2,J_3T_3}=-\frac
13\frac {i}{\Theta} \varepsilon_{cb3}\delta_{a3} .
\label{Eq21} \end{equation}
Using that both the axial  coupling constant and the isovector
magnetic moment are related to the corresponding form factors at $q^2=0 $
finally we get for the $1/N_c$ rotational corrections:
\begin{equation}
g_A(\Omega^1)=\frac {N_c}{9}\frac {i}{2\Theta} \sum_{\scriptstyle n >
val\atop {\scriptstyle m\leq val}}\frac
1{\epsilon_n-\epsilon_m}\bra{m}\tau_a\ket{n}
\bra{n}[\bm{\sigma}\times\bm{\tau}]_a\ket{m}
\label{Eq22} \end{equation}
and
\begin{equation}
\mu^{IV}(\Omega^1)=\frac {N_c}9\frac {i}{2\Theta} \sum_{\scriptstyle
n > val\atop {\scriptstyle m\leq val}}\frac
1{\epsilon_n-\epsilon_m}\bra{m}\tau_a\ket{n}
\bra{n}\gamma_5[[\bm{\sigma}\times\bm{x}]\times\bm{\tau}]_a\ket{n}
\label{Eq23} \end{equation}
In both eqns.(\ref{Eq22}) and (\ref{Eq23}), a summation over $a$ is
assumed. As should be expected both expressions have similar structure with
transitions  from occupied to non-occupied levels and back. They
include also an essential non-zero contribution from the Dirac sea. In
contrast to the leading order the above expressions are finite and one
does not need to regularize them.

The parameters of the model are fixed in the meson sector to reproduce
$\fpi=93$ MeV and $\mpi=139.6$ MeV. Similar to other
works~\cite{Reinhardt89W,TMeissner89,Wakamatsu91}
we use a numerical procedure based on the method of Kahana and
Ripka~\cite{Ripka84}. The results for $g_A$ and $\mu^{IV}$ up to
the first order terms in $\Omega$, are
presented in Fig.\ref{Figr2} and Fig.\ref{Figr3} as a function of the
constituent quark mass $M$. As can be seen in leading order ($\Omega^0$)
they are almost independent of the constituent quark mass and the
valence contribution is dominant. In the next to leading order
($\Omega^1$) the valence and Dirac sea contributions show much stronger
and quite different mass dependence: with increasing constituent mass
$M$ the valence part gets reduced whereas the Dirac sea part increases
and becomes dominant. However, their sum shows almost no dependence on
$M$. The result of Wakamatsu and Watabe for $g_A$ (labeled as W\&W in
Fig.\ref{Figr2}) deviates from the present numbers.
This is due to the fact that in their scheme the contribution of
the Dirac sea to $g_A$ in the linear order in $\Omega$ vanishes exactly. In
the present calculations for both quantities the enhancement due to the
$1/N_c$ rotational corrections improves considerably the agreement with
experiment. In the case
of $g_A$ the experimental value is almost exactly reproduced and the
inclusion of next order corrections will perhaps even overestimate it.
In contrast to $g_A$ for the isovector magnetic moment $\mu^{IV}$ we are
still below the experimental value by 25 \%.
It is interesting to notice that the
enhancement for both quantities due to the $1/N_c$ rotational corrections
is very close to the estimate $\frac {N_c+2}{N_c}$~\cite{Blaizot88}.

To conclude, we have evaluated the axial  coupling constant $g_A$ and
the isovector magnetic moment $\mu^{IV}$ in the Nambu -- Jona-Lasinio
model in the
next to leading order in the semiclassical quantization scheme. The $1/N_c$
rotational corrections are large enough to resolve the problem of strong
underestimation of these two quantities in the leading order. In particular,
$g_A$ is almost exactly reproduced. However, such large linear order
corrections imply that in order to control the
perturbation series in $\Omega$ the next order ($1/N_c^2$) corrections
should be investigated as well.

\begin{figure}
\vskip2cm
\caption{Diagrams corresponding to the expansion in $\Omega$ of the current
matrix element:
a) the valence contribution and b) the Dirac sea contribution.}
\label{Figr1}
\end{figure}
\begin{figure}
\caption{Axial vector coupling constant $g_A$ evaluated up to the linear
order in $\Omega$ as a function of the constituent quark mass $M$.}
\label{Figr2}
\end{figure}
\begin{figure}
\caption{Isovector nucleon magnetic moment $\mu^{IV}$ evaluated up to the
linear order in $\Omega$ as a function of the constituent quark mass $M$.}
\label{Figr3}
\end{figure}

\end{document}